\documentclass[12pt]{article}
\usepackage[dvips]{graphicx}
\usepackage{color}
\usepackage{amsfonts,amssymb}
\begin{document}

\vspace*{1cm}
\begin{center}
{\Large \bf The WA56 Experiment at CERN: How Does It Look\\[1ex]
in the Unified Picture for Hadron Spectra?}\\

\vspace{4mm}

{\large A. A. Arkhipov\footnote{e-mail: arkhipov@mx.ihep.su}\\
{\it State Research Center ``Institute for High Energy Physics" \\
 142281 Protvino, Moscow Region, Russia}}\\
\end{center}

\vspace{2mm}
\begin{abstract}
In this note we demonstrate how the experimental material taken from
WA56 Experiment at CERN looks in the recently developed  unified
picture for hadron spectra. Our analysis shows that baryonium states
observed in the WA56 Experiment are the states living in the
corresponding KK tower built in according to the earlier established
general, physical law.
\end{abstract}

\section*{}

There were several experiments made with the CERN OMEGA spectrometer.
The WA56 Experiment was one of them especially performed to study the
baryon exchange processes among hadronic reactions. Perhaps, it
should be reminded about an old and long-standing problem of narrow
$pp$ (dibaryon) states and $p\bar p$ (baryonium) states. In old, up
to the end of the eighties, issues of PDG booklets \cite{1} we can
find a separate parts with a compilation of these states. However,
that compilation has been rejected by PDG later on probably because
several experiments made in the beginning and the middle of the
eighties had failed to search for these states. That is why, it might
seem the interest to these states disappeared. However, this is not
the case. In fact, up to now, narrow dibaryon and baryonium states in
two-nucleon system have always attracted incessant interest from both
theorists and experimenters. The physical origin of such narrow
states is high interest because it has fundamental importance related
to the nature of fundamental nucleon-nucleon forces.

We start with a short review of some known experimental studies
devoted entirely to search for narrow $p\bar p$ states. In paper
\cite{2} an evidence for two narrow $p\bar p$ resonances at 2020 MeV
and 2200 MeV had been reported. From the study of the reaction
$\pi^-p\rightarrow p_f\pi^-p\bar p$ using a fast proton trigger
device in the CERN OMEGA spectrometer, these two narrow $p\bar p$
states were observed mainly in association with a $\Delta^0$(1232)
and $N^*$(1520) production in top vertex of the baryon exchange
diagram (see Fig. 1a). The statistical significance of each peak was
greater than 6 standard deviations. Masses and natural widths of
these resonances were respectively $M_1$ = 2020 $\pm$ 3 MeV,
$\Gamma_1$ = 24 $\pm$ 12 MeV and $M_2$ = 2204 $\pm$ 5 MeV, $\Gamma_2$
= 16 $^{+20}_{-16}$ MeV. The upper limits to the production cross
sections were estimated about $\sim$10-30 nb. These results were part
of a general experiment made with the CERN OMEGA spectrometer and
came from a special experiment devoted to a general study of the
baryon exchange processes in the reaction chain
\begin{equation}
\pi^- + p \to \Delta^{0}(1232)[N^{*}(1520)] +  M^0 , \quad\
\Delta^{0}(1232)[N^{*}(1520)]\to p\pi^- , \quad M^0\to p\bar p.
\label{barexch}
\end{equation}
The experiment was done with a $\pi^-$ beam at an incident momentum 9
and 12 GeV colliding with a hydrogen target. The trigger required a
fast proton emitted forward with a momentum greater than
0.5$p_{\,beam}$. A complete description of this experiment and the
details of analysis can be found in original paper \cite{2} and
references therein.

The results of a search for narrow $p\bar p$ states produced in the
baryon exchange reaction with a chain
\begin{equation}
\pi^+ + p \to \Delta_f^{++}(1232) +  X^0 , \quad\
\Delta_f^{++}(1232)\to p_f\pi^+ , \quad X^0\to p\bar p
\label{barexchBNL}
\end{equation}
at $p_{\,beam}^{\pi^+}$ = 9.8  GeV have been reported in Ref.
\cite{3}. As pointed out, this channel provided an enhanced
sensitivity to such states compared with reaction (\ref{barexch})
studied in Ref. \cite{2}. The multiparticle spectrometer (MPS) at the
Brookhaven National Laboratory alternating-gradient synchrotron was
used to detect all four charged tracks from reaction
(\ref{barexchBNL}). No evidence for narrow $p\bar p$ states at the
20--30 nb level in the mass range 1.9--2.3 GeV was found. It was also
emphasized that, assuming a nucleon-exchange production mechanism,
the states which have been reported in Ref. \cite{2} would appear as
$>$ 5 standard- deviation peaks in the recorded data sample with the
trigger required a forward particle with momentum $\geq$ 5.5 GeV.

The $p\bar p$ mass spectrum  was also measured in another experiment
\cite{4} at the Brookhaven National Laboratory alternating-gradient
synchrotron where the inclusive reaction $\pi^- + p(or\, C)\to p\bar
p + X^0$ had been studied. No statistically significant enhancements
in the data in the $p\bar p$ mass range from 2000 to 2400 MeV were
observed.

A search was made for baryonium production in $p\bar p$ and $K^-p$
interactions at 12 GeV in the experiment \cite{5} which was performed
using the OMEGA spectrometer at CERN, exposed to the separated beam
consisting of approximately 45\% antiprotons, 15\% negative kaons and
40\% negative pions. No significant structures in the mass spectra of
$p\bar p$, $p\bar p\pi^-$ and $\bar p\Lambda^0$ systems were seen.
The upper limit at the 99.5\% confidence level for production of
narrow states in the $p\bar p$ system in the mass range 2.0--2.2 GeV
was estimated about $\sim$40 nb.

It should also be mentioned that the negative results on backward
production, via baryon exchange, of exotic non-strange mesons were
presented in article \cite{6}. The reactions $\pi^-p\to
p_{forward}X^-$ and $\pi^-n\to p_{forward}X^{--}$ have been studied
with a 12 GeV $\pi^-$ beam in the OMEGA spectrometer at CERN. No
resonant peak in $X\to p\bar p\pi^-,\, p\bar p\pi^-\pi^-,\, p\bar
p\pi^-\pi^0,\,$ $\pi^+\pi^-\pi^-\pi^-,\, \pi^+\pi^-\pi^-\pi^0$ has
been seen. The upper limits obtained on cross sections for exotic
meson production $X\to N\bar N\pi,\, N\bar N\pi\pi,\, 4\pi$ were
lower than the $\rho^-$ backward production cross section in the $\pi
p\to p\rho^-$ reaction. As pointed out, the sensitivity of this
experiment increased by an order of amplitude compared to earlier
published experiments in the search for exotics produced via baryon
exchange.

The WA56 experiment at CERN \cite{7} was also one of the experiments
made with the OMEGA spectrometer specially designed to select the
baryon exchange processes. It was a long-range aim of the WA56
experiment to confirm the narrow $p\bar p$ states of masses 2020 MeV
and 2204 MeV reported in paper \cite{2}. However, the WA56 experiment
revealed the lack of these states at the level of production cross
sections smaller by an order of magnitude than those estimated in
Ref. \cite{2}. In fact, the results  of a search for narrow $p\bar p$
states produced backwards in the baryon exchange reactions $\pi^-p\to
\Delta_f^0(1232)\bar pp$, $\pi^-p\to N_f^0(1520)\bar pp$ at 12 GeV
and $\pi^+p\to \Delta_f^{++}(1232)\bar pp$ at 20 GeV have been
reported. No structures of statistical significance exceeding three
standard deviations have been found in the $p\bar p$ mass spectra.
The cross section limits obtained were three to five times lower than
the cross sections of Ref. \cite{2} depending on the channel.

Quite a new analysis of the WA56 experimental data was performed in
the nineties \cite{8} with a chief aim to study the channels of the
baryon exchange reactions which were not taken into consideration
before. In fact, the baryon exchange reactions  have been studied
where one particle was either undetected or incompletely
reconstructed but the $p\bar p$ system was produced in central region
(see Fig. 1c). Namely, the following channels have been considered
\begin{equation}
\pi^+ + p \to p_f +  X^0 + \pi_s^+, \quad X^0\to p\bar p,
\label{barexch1}
\end{equation}
\begin{equation}
\pi^+ + p \to p_f + \pi^+ + X^0 + \pi_s^0, \quad X^0\to p\bar p,
\label{barexch2}
\end{equation}
\begin{equation}
\pi^+ + p \to p_f + \pi^+ + \pi^+ + X^0 + \pi_s^-, \quad X^0\to p\bar
p, \label{barexch3}
\end{equation}
at 20 GeV and
\begin{equation}
\pi^- + p \to p_f +  X^0 + \pi_s^-, \quad X^0\to p\bar p,
\label{barexch4}
\end{equation}
\begin{equation}
\pi^- + p \to p_f + \pi^- + X^0 + \pi_s^0, \quad X^0\to p\bar p,
\label{barexch5}
\end{equation}
at 12 GeV, where $p_f$ was an identified fast proton with the
momentum greater than half the beam momentum, and the slow pions
($\pi_s^0,\pi_s^\pm$) went undetected by the OMEGA spectrometer but
were reconstructed by the corresponding kinematic fits; see, however,
the details in original paper \cite{8}. As pointed out, the very good
momentum resolution available from OMEGA tracks measurements allowed
a clear separation and identification of one missing pion channel in
the data. That investigation was motivated in the main by previous
analysis \cite{9} of the WA56 experimental data on the central
production of $\rho^0$, $f_2$ and $\rho_3^0$ mesons in the baryon
exchange reaction

\begin{equation}
\pi^+ + p \to p_f +  M^0 + \pi_s^+, \quad M^0\to \pi^+\pi^-,
\label{barexch6}
\end{equation}
at 20 GeV (see Fig. 1b) where the similar experimental approach was
used for the first time. A clear signal of a narrow $p\bar p$ state
with a mass of 2.02 GeV and a width less 10 MeV was seen in all of
the channels (\ref{barexch1}-\ref{barexch5}) in a restricted
kinematic region close to the central one. From the upper limits on
mesonic ($\pi^+\pi^-$, $\pi^+\pi^-\pi^0$, $2\pi^+2\pi^-$, $K^+K^-$)
decay modes of the observed state it was found that this state was
not noticeably coupled to mesons. That is why, it was claimed that
this 2.02 GeV $p\bar p$ state coupled strongly to baryons and
decoupled from mesons might be a baryonium candidate.

The combined $p\bar p$ mass spectrum for the events in $\pi^+p$
(channels (\ref{barexch1},\ref{barexch2})) and $\pi^-p$ (channels
(\ref{barexch4},\ref{barexch5})) exposures, plotted with the 5 MeV
width bins, is shown in Figure 2 extracted from original paper
\cite{8}. As seen in Fig.~2, the clear signal is visible at a mass of
2.02 GeV. The statistical significance of this peak was estimated to
be exceeded 5 standard deviations. It should also be pointed out that
no statistically significant peak at a mass of 2.20 GeV, reported
earlier in Ref. \cite{2}, was observed in the data. However, at the
same time, the mass fit results for the 2.02 GeV $p\bar p$ state
accurately collected in Table 1 of Ref. \cite{8} turned out in a good
agreement with the similar results of Ref. \cite{2}.

Of course, the question arises: could the contradicted experimental
results on (non)ob\-servation of narrow $p\bar p$ states be really
compatible? In Ref. \cite{8} it has been presented a shining example
of that how the discrepancies could be explained. First of all, it
has been pointed out that all experimental results on the production
cross sections depend on the model which has been used for reaction
mechanism in order to extract the experimental acceptances. For
example, the experimental acceptances were calculated in Ref.
\cite{2}, using mechanism of backward production of the 2.02 GeV
$p\bar p$ state. For a proper comparison of the results the authors
of Ref. \cite{8} recalculated the acceptance of the experimental
setup \cite{2} assuming their central production mechanism. As a
result with the revised acceptance calculation, the production cross
section of the 2.02 GeV $p\bar p$ state was found to be in experiment
\cite{2} such as obtained in experiment \cite{8} with a good
agreement. The same has turned out true in a comparison of the
results from the experiments \cite{7} and \cite{8}. In other words,
the authors of Ref. \cite{8} have simply explained how to get an
agreement with the experimental results on the production of the 2.02
GeV $p\bar p$ state reported in \cite{2,7} and their own experiment
\cite{8}. An explanation of the absence of the 2.02 $p\bar p$ state
in other experiments (see e.g. \cite{3,4,5}) has also given in Ref.
\cite{8}, and we refer the interested reader to the original paper.

Concerning formation experiments it was shown in Ref. \cite{8} that
the experimental cross section of the (formation) process $N\bar N
\rightarrow R \rightarrow N\bar N$ at the resonance peak crucially
depends on the ratio of the resonance width to the experimental
resolution
\begin{equation}
\sigma^{N\bar N \rightarrow N\bar
N}_{exp}(\sqrt{s}=\sqrt{s_R})\vert_{M_R=2.02}\simeq
2.25(2J+1)\beta^2_{N\bar
N}\frac{\Gamma}{2\Delta}\arctan\frac{2\Delta}{\Gamma}\,\mbox{mb},
\label{sigmaexp}
\end{equation}
where $J$, $\beta_{N\bar N}$, $\Gamma$ are spin, elasticity
($\beta_{N\bar N}=1$ for a pure baryonium decaying into $N\bar N$
only), width of the resonance, $\Delta$ is experimental resolution.
The experimental restriction on the width of the 2.02 GeV $p\bar p$
state obtained in Ref. \cite{8} ($\Gamma\leq 10$)MeV should be taken
into account to understand the discrepancies with the formation
experiments where no signal of the 2.02 GeV $p\bar p$ state was
found.

It should also be mentioned a remark in Ref. \cite{8} on the central
production mechanism priority over the backward production one. In
fact, this peculiarity was established earlier in the study of
central production of ordinary mesons \cite{9} where the enhancement
factor did not exceed 4, but for the 2.02 GeV $p\bar p$ state
production this factor turned out at least 20. To explain such
peculiarity it may be assumed that the observed narrow $p\bar p$
state is coupled with the $\Delta\bar\Delta$ system much stronger
than with the $N\bar N$ one. In that case a process of backward
production, where double $\Delta$ exchange does not work, is
naturally suppressed, and the dominated process is given by diagram
1c shown in Fig. 1. Here the narrowness of the observed $p\bar p$
state might also be explained as follows \cite{8}: this state being
produced in the collision of virtual $\Delta$ and $\bar\Delta$ cannot
decay into the real pair $\Delta\bar\Delta$ because this decay
channel is forbidden by phase space, and only suppressed $p\bar p$
channel is open.

Bear all of that in mind, we would like to emphasize that careful
analysis performed in Ref. \cite{8} might be served as an excellent
introduction to the recent widely discussed question: why the
$\Theta$ baryon states observed in a more than 10 experiments are not
seen in the others?

In Ref. \cite{10}, where some of our previous studies were partially
summarized, it has been claimed that existence of the extra
dimensions in the spirit of Kaluza and Klein together with some novel
dynamical ideas may provide new conceptual issues for the global
solution of the spectral problem in hadron physics to build up a
unified picture for hadron spectra. Earlier we have applied these
ideas to analyze the nucleon-nucleon dynamics at very low energies
\cite{11}. Really, we have found that simple formula for KK
excitations provided by Kaluza-Klein approach accurately described
the experimentally observed irregularities in the mass spectrum of
$pp$ and $p\bar p$ systems. The result of our analysis is presented
in Table 1 extracted from Ref. \cite{11} (see also the references
therein where the experimental data have been extracted from). As is
seen from Table 1, the nucleon-nucleon dynamics at low energies
reveals quite a remarkable development of Kaluza-Klein picture.
Moreover, $M_n^{pp}=M_n^{p\bar p}$ is predicted by Kaluza-Klein
scenario, and Table 1 contains an experimental confirmation of this
fact as well.

Of course it was intriguing for us to perform the spectral analysis
of the experimental material from the WA56 Experiment at CERN in
order to learn how does it look in the unified picture for hadron
spectra. To attain this goal the calculated spectral lines taken from
Table 1 have been plotted in Fig.~3 together with the $p\bar p$ mass
spectrum presented in Ref. \cite{8}. As is seen, the experimentally
observed 2.02 GeV $p\bar p$ state just lives in the $M_{9}^{p\bar
p}$(2019.66\,MeV)-storey of KK tower for the $p\bar p$ system.  What
is more important, a strong correlation of the spectral lines with
the other peaks on the histogram is also clear seen in Fig.~3. In our
opinion, that correlation is not an accidental coincidence, it
displays the existence of the states observed in other experiments.
In fact, the different experiments reflect only partial fragments of
the whole unified picture.

Our conservative estimate for the widths of KK excitations looks like
\begin{equation}\label{width}
\Gamma_n \sim \frac{\alpha}{2}\cdot\frac{n}{R}\sim 0.4\cdot n\,
\mbox{MeV},
\end{equation}
where $n$ is the number of KK excitation, and $\alpha \sim 0.02$,
$R^{-1}=41.48\,\mbox{MeV}$ are known from our previous studies
\cite{11}. This gives $\Gamma_{9}\sim 3.6$ MeV which does not
contradict to the experimental estimate.

At last, we also predicted the narrow charged $p\bar n$ and $\bar pn$
states with the similar mass: $M_9^{p\bar n}=M_9^{\bar pn}= 2020.84$
MeV (see Table 2 in Ref. \cite{11}). An evidence for the existence of
such states has been reported in Ref. \cite{12}, where the authors
claimed that these states were observed in the reactions $\bar
pp\rightarrow p_f\bar n\pi^+\pi^-\pi^-$ at 6 GeV and $\bar
pp\rightarrow \pi_f^+\bar pn\pi^+\pi^-$ at 9 GeV in a triggered
bubble chamber experiment at the SLAC Hybrid Facility. Clearly, this
experimental observation is an additional argument in favour of the
Kaluza-Klein picture.

In summary, the results of the WA56 Experiment at CERN, with account
of analysis made in \cite{8}, are naturally incorporated in the
recently developed unified picture for hadron spectra. Our analysis
shows that the experimentally observed narrow baryonium states are
the states living in the corresponding KK tower built in according to
the earlier established general, physical law. We hope that new
experiments will appear in the near future to enrich our
understanding of the nucleon-nucleon dynamics. We also share that a
call for the future hadronic experiments with high resolution and
sensitivity has clearly to be supported.

\newpage
\vspace*{2cm}
\begin{center}
Table 1. Kaluza-Klein tower of KK excitations for $pp(p\bar p)$
system and experimental data.

\vspace{5mm} \hbox to \hsize {\hss
\begin{tabular}{|c|c|l|l||c|c|l|l|}   \hline \vspace{-4.5mm}
 & & & & & & & \\
 n & $M_n^{pp}$\,MeV & $M_{exp}^{pp}$\,MeV &
$M_{exp}^{p\bar p}$\,MeV & n & $M_n^{pp}$\,MeV & $M_{exp}^{pp}$\,MeV
& $M_{exp}^{p\bar p}$\,MeV \\ \hline 1 & 1878.38 & 1877.5 $\pm$ 0.5 &
1873 $\pm$ 2.5 & 15 & 2251.68 & 2240 $\pm$ 5 & 2250 $\pm$ 15
\\ \hline
2 & 1883.87 & 1886 $\pm$ 1 & 1870 $\pm$ 10 & 16 & 2298.57 & 2282
$\pm$ 4 & 2300 $\pm$ 20  \\ \hline 3 & 1892.98 & 1898 $\pm$ 1 & 1897
$\pm$ 1 & 17 & 2347.45 & 2350 & 2340 $\pm$ 40
\\ \hline
4 & 1905.66 & 1904 $\pm$ 2 & 1910 $\pm$ 30  & 18 & 2398.21 &  & 2380
$\pm$ 10
\\ \hline
5 & 1921.84 & 1916 $\pm$ 2 & $\sim $ 1920 &  19 & 2450.73 &   & 2450
$\pm$ 10   \\ \cline{5-8}
  &         & 1926 $\pm$ 2 &   & 20 & 2504.90 &   & $\sim$
  2500    \\ \hline
  &         & 1937 $\pm$ 2 & 1939 $\pm$ 2 & 21 & 2560.61 &
  &  \\ \cline{5-8}
6 & 1941.44 & 1942 $\pm$ 2 & 1940 $\pm$ 1 & 22 & 2617.76 &  & $\sim$
2620
  \\ \cline{5-8}
  &         & $\sim$1945  & 1942 $\pm$ 5 & 23 & 2676.27 &  & \\ \hline
7 & 1964.35 & 1965 $\pm$ 2 & 1968 & 24 & 2736.04 & 2735 & 2710 $\pm$
20  \\ \cline{5-8}
  &         & 1969 $\pm$ 2 & 1960 $\pm$ 15 & 25 & 2796.99 &  &
  \\ \hline
8 & 1990.46 & 1980 $\pm$ 2 & $1990^{\,+15}_{\,-30}$ & 26 & 2859.05 &
& 2850 $\pm$ 5 \\ \cline{5-8}
  &         & 1999 $\pm$ 2 &  &  27 & 2922.15 &  &
\\ \hline
9 & 2019.63 & 2017 $\pm$ 3 & 2020 $\pm$ 3 & 28 & 2986.22 & & $\mathbf{2984 \pm 2.1 \pm 1.0}$  \\
\hline 10 & 2051.75 & 2046 $\pm$ 3 & 2040 $\pm$ 40  & 29 & 3051.20 &
&
   \\ \cline{5-8}
   &         & $\sim$2050 & 2060 $\pm$ 20 &  30 & 3117.04 &  &  \\ \hline
11 & 2086.68 & 2087 $\pm$ 3 & 2080 $\pm$ 10 &  31 & 3183.67 &  &  \\
\cline{5-8}
   &         &            & 2090 $\pm$ 20 & 32 & 3251.06 &  &  \\ \hline
   &         & $\sim$2122 & 2105 $\pm$ 15 & 33 & 3319.15 &  & \\ \cline{5-8}
12 & 2124.27 & 2121 $\pm$ 3 & 2110 $\pm$ 10 & 34 & 3387.90 &  & 3370
$\pm$ 10 \\ \cline{5-8}
  &         & 2129 $\pm$ 5 & 2140 $\pm$ 30 & 35 & 3457.28
&  & \\ \hline 13 & 2164.39 & $\sim$2150 & 2165 $\pm$ 45 & 36 &
3527.25 &   & $h_c(1P)(3526)$ \\ \cline{5-8}
   &         & 2172 $\pm$ 5 & 2180 $\pm$ 10 & 37 & 3597.77 &
   & 3600
$\pm$ 20         \\ \hline 14 & 2206.91 & 2192 $\pm$ 3 & 2207 $\pm$
13 & 38 & 3668.81 &  &  \\ \hline
\end{tabular}
\hss}

\end{center}

\newpage
\vspace*{6cm}

\begin{figure}[htb]
\begin{center}
\includegraphics[width=\textwidth]{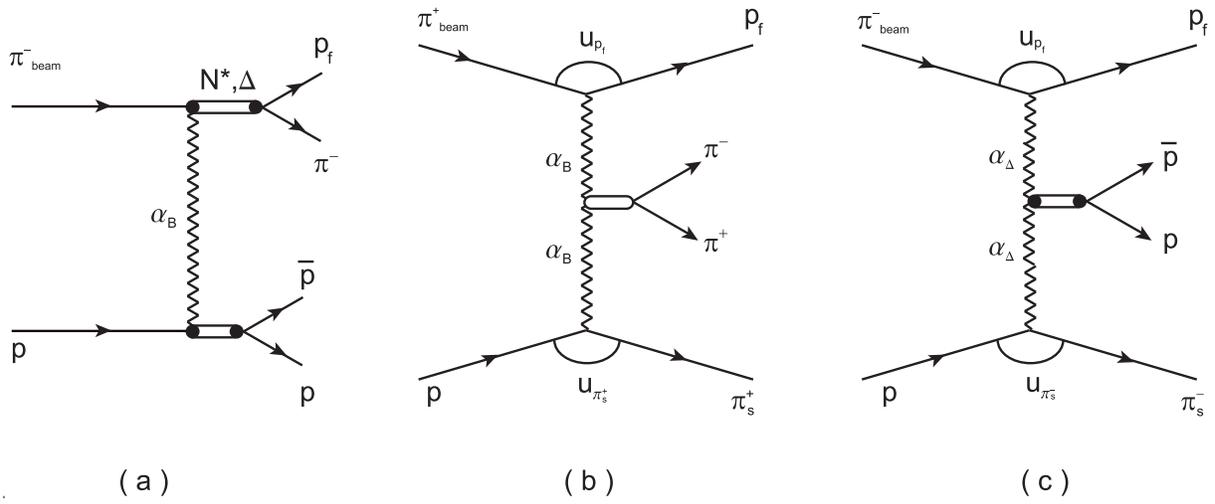}\label{fig1}
\end{center}
\caption{The diagrams from Ref. \cite{8} which display: the backward
production of the $p\bar p$ state in  $\pi^{-}p$ interactions (a),
the central production of a meson in $\pi^{+}p$ interactions (b), the
central production of the $p\bar p$ state in  $\pi^{-}p$ interactions
via double $\Delta$ exchange (c).}
\end{figure}

\newpage
\vspace*{6cm}

\begin{figure}[htb]
\begin{center}
\includegraphics[width=\textwidth]{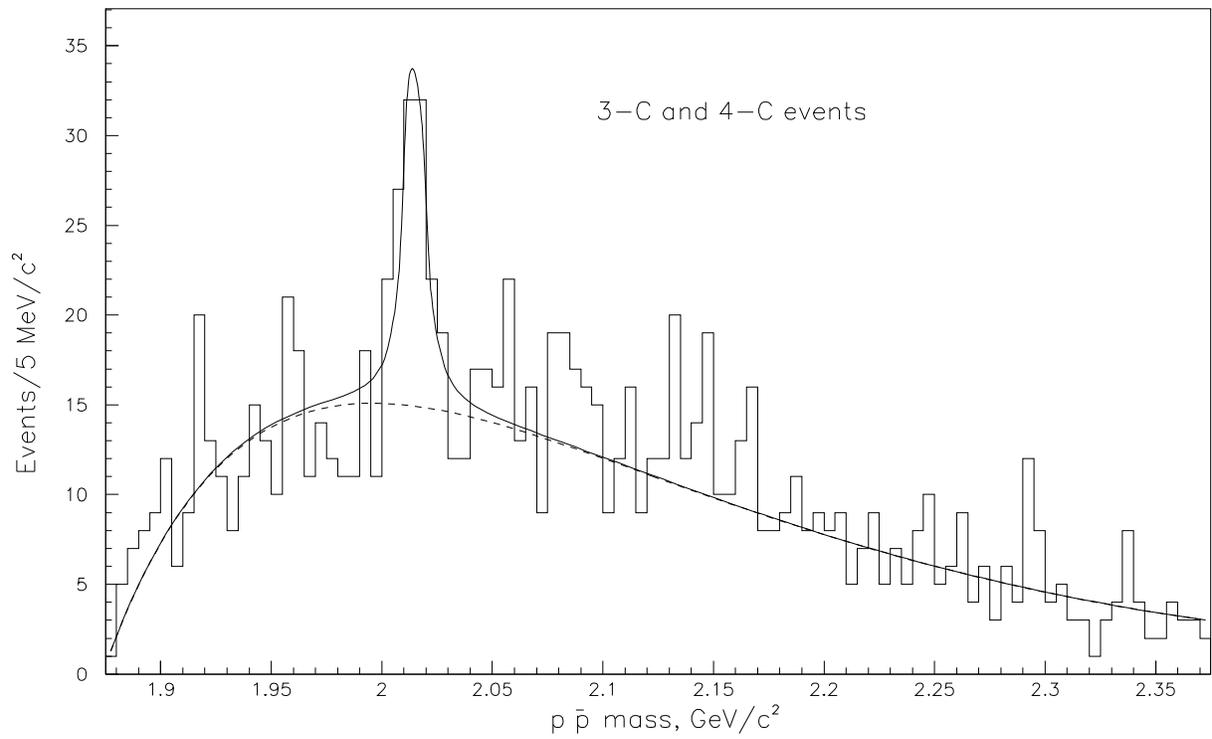}\label{fig2}
\end{center}
\vspace*{-1.5cm} \caption{Combined $p\bar p$ mass spectrum for the
events of reactions (\ref{barexch1},\ref{barexch4}) and
(\ref{barexch2},\ref{barexch5}) presented in Ref. \cite{8}.}
\end{figure}

\newpage
\vspace*{6cm}

\begin{figure}[htb]
\begin{center}
\includegraphics[width=\textwidth]{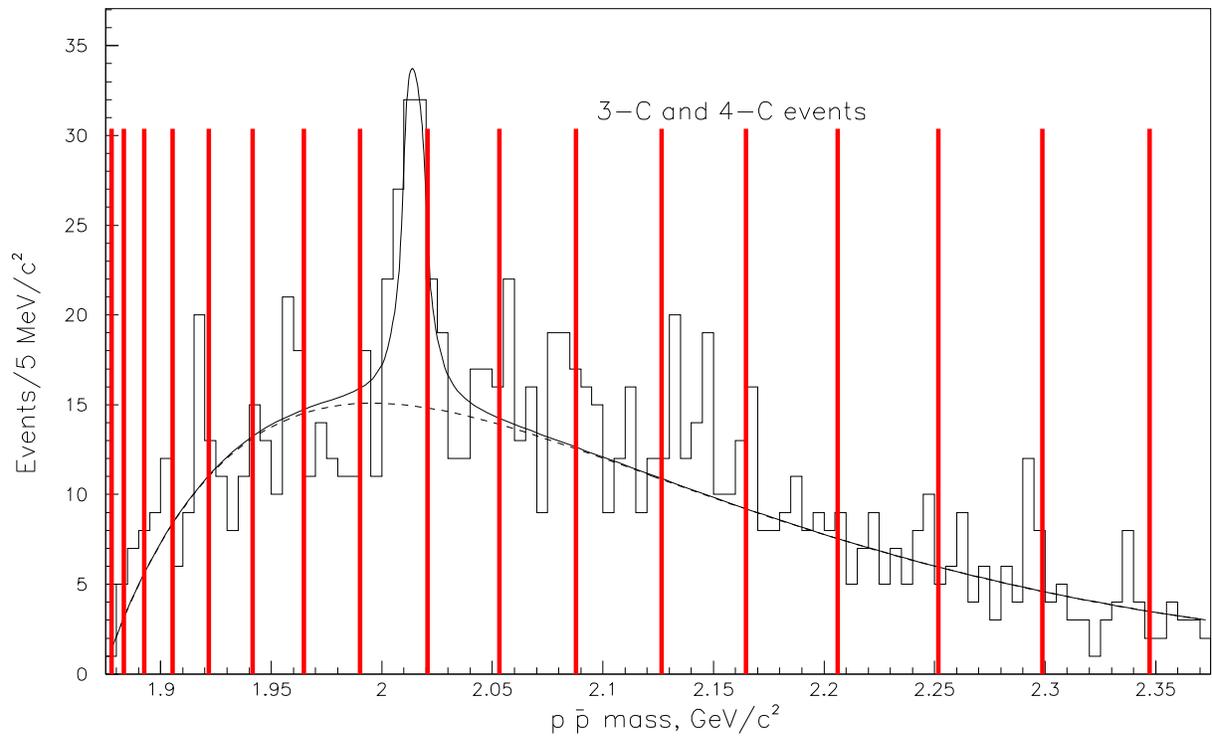}\label{fig3}
\end{center}
\vspace*{-0.5cm}
\caption{The same as Fig.~2 but with the vertical
(spectral) lines corresponding to KK tower for $p\bar p$ system; see
Table 1.}
\end{figure}

\end{document}